\documentclass{article}
\usepackage{spconf,amsmath,graphicx}

\title{Refining Automatic Speech Recognition System for older adults}
\name{Liu Chen, Meysam Asgari}

\address{Center for Spoken Language Understanding,
Oregon Health \& Science University,  
Portland, Oregon, USA\\}
\begin{document}
%
\maketitle
\begin{abstract}
Building a high quality automatic speech recognition (ASR) system with limited training data has been a challenging task particularly for a narrow target population. Open-sourced ASR systems, trained on sufficient data from adults, are susceptible on seniors’ speech due to acoustic mismatch between adults and seniors. With 12 hours of training data, we attempt to develop an ASR system for socially isolated seniors (80+ years old) with possible cognitive impairments. We experimentally identify that ASR for the adult population performs poorly on our target population and transfer learning (TL) can boost the system’s performance. Standing on the fundamental idea of TL, tuning model parameters, we further improve the system by leveraging an attention mechanism to utilize the model’s intermediate information. Our approach achieves $1.58\%$ absolute improvements over the TL model.  
\end{abstract}
\section{Introduction}
\label{sec:intro}
Recently, in cognitive research, analyzing everyday conversation has received increasing attention as it opens a window toward to individuals’ personal world and could potentially reveal their cognitive and behavioral characteristics~\cite{khodabakhsh2015evaluation}. This analysis relies on high-fidelity transcription which is labor intensive and costly. A high-quality ASR system could potentially serve as an alternative solution to facilitate the analyzing process. The ASR systems have been widely used in  medical applications such as  clinical documentation~\cite{zhou2018analysis} and healthcare systems~\cite{ismail2020development}.
The ASR systems have also been adopted in medical researches, such as cognitive tests in Alzheimer's research~\cite{konig2015automatic}. 

Nowadays, deploying deep neural networks (DNN) in key components of the ASR engine, the acoustic and language models, has significantly boosted the performance. On the other hand, ASR systems inherit DNN’s hunger for target-domain data~\cite{sun2017revisiting} and the susceptibility to domain mismatch. Although training a domain-specific model with sufficient data is ideal, collecting training data is a challenging task in the medical field especially when facing a narrow target population~\cite{holden2015data}.

This challenge can be caused by multiple factors, such as target population size, financial limitations or policy concerns. Current publicly available ASR systems, which are trained on large amounts of data from adults, perform well on their training domain. However, their performance degrades in clinical applications for seniors as the acoustic characteristics of the target speakers deviate from those used in training examples. This is a crucial limitation in our proposed study as the age range of our participants (seniors above 80 years old) imposes a strong acoustic mismatch leading to an inaccurate recognition. Seniors' vocal characteristics are different from adults. Age-related speech deterioration begins around 60 years old~\cite{minifie1994introduction} resulting in significantly different voice characteristics in comparison to the younger generation~\cite{linville2002source}. Enriching seniors' training dataset with adult’s recordings cannot ease the data restriction. Moreover, plausible influence of impaired cognitive functioning  on acoustic features of MCI subjects~\cite{rodgers2013influence} may serve as an additional source of acoustical mismatch. 

Our training dataset contains 12 hours of transcribed recordings collected from above 80 years old seniors with possible cognitive impairments. To train a end-to-end ASR system, we propose using transfer learning (TL) to address the data limitation. Based on the idea of transfer learning, we design a conditional-independent attention mechanism to leverage the intermediate outputs from the base model. Our method gains a lower word error rate (WER) score than weight transfer learning (WTL). 

The remainder of this paper is organized as follows. Section~\ref{sec: background} presents backgrounds of transfer learning and the attention mechanism. Section~\ref{sec: uncondi_atten} describes the conditional-independent attention mechanism and its application. Section~\ref{sec: data} present the detail of our experimental data. Section~\ref{sec: exp_atten} describes the experimental setup and presents testing results.

\section{Background}
\label{sec: background}

\subsection{Transfer Learning}
\label{sec: intro_transf}
TL is a type of adaptation widely used in data-scarce scenarios aiming to reuse the learned knowledge of a base DNN model on a related target domain. The fundamental idea is tuning the weights of a pre-trained model. WTL and hidden linear TL (HLTL) are two well-studied ones. WTL initializes the target model with a pre-trained base model where both models share the exact same DNN architecture. Then, tune the entire or a portion of the DNN with a small learning rate~\cite{ shivakumar2018transfer}. 
HLTL builds upon a hypothesis where similar tasks share common low-level features. It considers the base model as part of the target model by using the base model as a feature extractor that makes the first process on the input data~\cite{li2010comparison}. Other than these well-studied methods, Factorized Hidden Layer (FHL) models the pre-trained parameters with a linear interpolation of a set of bases~\cite{samarakoon2016factorized,sim2018domain}. A key ingredient to a successful model adaptation is a well-trained base model from a similar task that has learned from a relatively large and yet diverse training samples~\cite{yosinski2014transferable}.

\subsection{Attention Mechanism}
\label{sec:atten_mechanism}
\begin{figure}[htb]
    \centering
    \includegraphics[width=8.5cm]{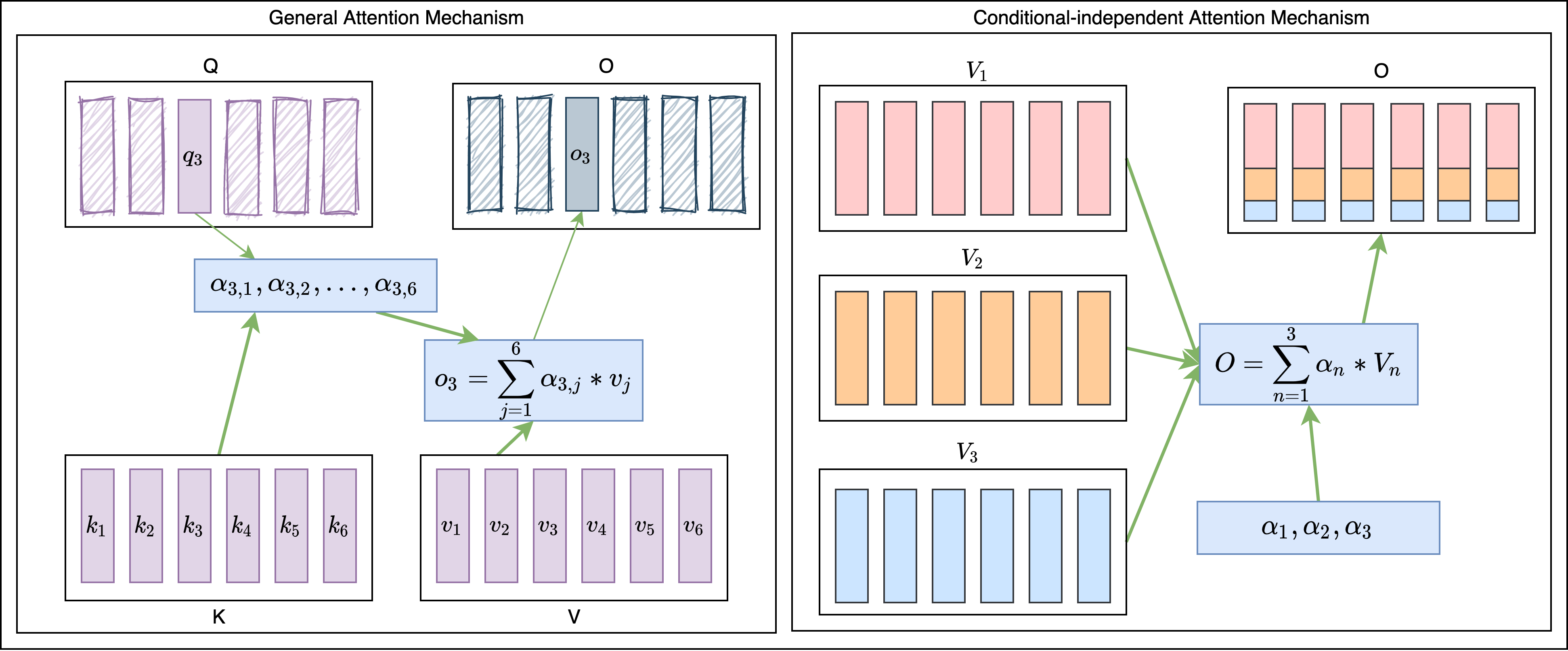}
    \caption{The left graph shows the backbone of the general attention mechanism. The right graph is the backbone of our conditional-independent attention mechanism.}
    \label{fig:vis_exp_atten}
\end{figure}

Researchers utilized the attention mechanism to manage long sequence memory without forgetting~\cite{graves2014neural}. Given an input vector, the mechanism aligns the input to each memory cell, and outputs the summarized memory. The following equations are the general operations of the mechanism:
\begin{align*}
score_{t,j} &= \langle \phi(q_t), \psi(k_j) \rangle\\
\alpha_{t,j} &= align(q_{t}, k_{j})\\
&= \frac{score_{t,j}}{\sum_{j'=1}^{length(V)}score_{t,j'}}\\
&= softmax(score_{t,*}) \\
o_{t} &= \sum_{j=1}^{length(V)}\alpha_{t,j} * v_{j}
\end{align*}   
where $q_t$ is the $t$th column of matrix $Q$. The vectors $k_j$ and $v_j$ are the $j$th columns in matrices $K$ and $V$, respectively. These two matrices represent the identity and the content of the memory. The symbols, $\phi$ and $\psi$, represent linear layers. The left graph in Figure~\ref{fig:vis_exp_atten} describes the general structure of this mechanism. The alignment is conditional to the input $k$. Although the conditional alignment shows its power in multiple researches~\cite{graves2014neural,luong2015effective}, a small dataset is not sufficient to train these additional layers in TL. In Section~\ref{sec: uncondi_atten}, we will introduce our conditional-independent attention mechanism which only requires a limited amount of additional parameters. 
 
\subsection{DeepSpeech2} 
\label{sec: ds2}
\begin{figure}[htb]
    \centering
    \includegraphics[width=\linewidth]{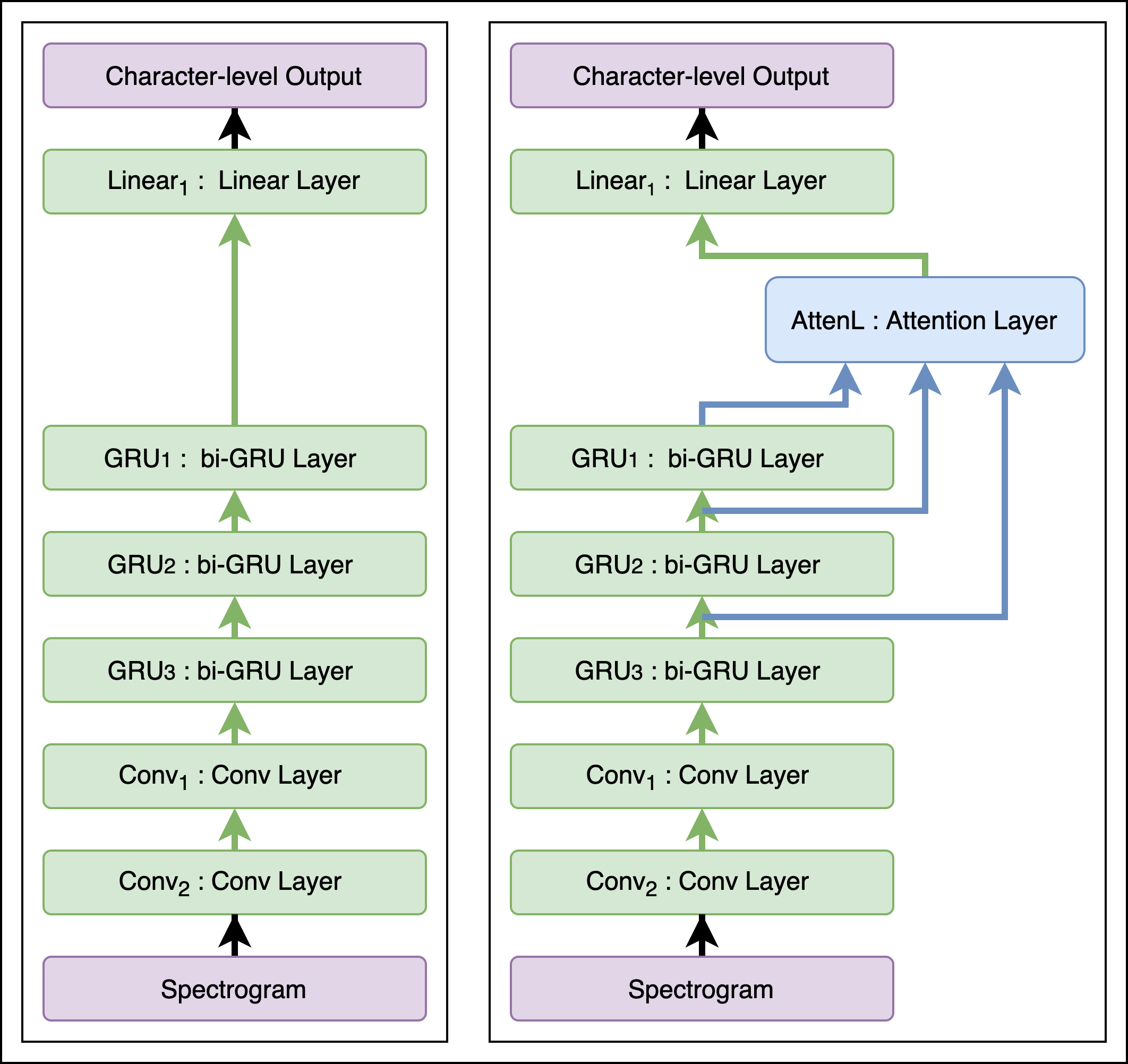}
    \caption{The left graph is the architecture of DS2 which is our base model. The right graph shows the modified DS2 architecture to leverage intermediate outputs. Each box contains the layer's nickname and type.}
    \label{fig:miniAttention}
\end{figure}

DS2~\cite{amodei2016deep} is an end-to-end ASR system that leverages a recurrent neural network (RNN) for modeling the spectrogram to the sequence of 26  English letters. This architecture utilized convolution (Conv) layers for low-level information processing and uses bidirectional gate recurrent unit (bi-GRU) layers for high-level processing. The output layer is a linear layer. In the decoding process, the DS2~\cite{amodei2016deep} utilizes a beam search approach~\cite{wiseman2016sequence} to search for the transcription with the highest probability based on the combination of probabilities from the DS2 model and a n-gram language model. The left graph of Fig~\ref{fig:miniAttention} is the architecture that we used in our experiments. 

\section{Conditional-independent Attention Mechanism}
\label{sec: uncondi_atten}
Based on the fundamental idea of TL, we  focus on leveraging intermediate information from the base model. FTL utilizes these information to interpolate trained model parameters in speaker adaptation~\cite{samarakoon2016factorized} and domain adaptation~\cite{sim2018domain}. We assume hidden output loss part of useful information which can be retrieved from intermediate outputs. Moreover, we also extend the use case of attention mechanism. We still consider it as a control unit. Instead of long-term memory management, we utilize the mechanism to summarize useful information for target domain.  The right graph in Fig~\ref{fig:vis_exp_atten} shows the backbone of our mechanism. Theoretically, we can use it in any network and the use case is not limited to TL. But, we focus our research on the data-scarce problem. The right graph in Fig~\ref{fig:miniAttention} is the modified DS2 architecture where we add an Attention layer (AttenL) to summarize bi-GRU layers’ outputs. Although all examples takes three inputs, the mechanism does not limit to this number. We modified the standard attention into the conditional-independent version: (1) the manual attention and (2) the learnable attention. 

\subsection{Manual Unconditional Attention}
 We remove the alignment function from the standard attention mechanism and manually assign the attention to the outputs of GRU layers based on our knowledge about the model. The following equation is the operations of manual attention:
\begin{align*}
    O &= \sum_{n=1}^{3}\alpha_{n} * V_{n}
\end{align*}
where $V_n$ is the output of GRU$_n$ and $\alpha_{n}$ is the attention that we manually assign to it. This format is the same as the weighted average of GRU outputs. This does not have additional parameters. Therefore it is ideal for evaluating the effectiveness of utilizing intermediate information. We name this mechanism as the manual attention mechanism (MAM).

\subsection{Learnable Attention}
To learn the attention from the target dataset, we apply a function to learn assigning attention in a conditional-independent fashion. The learnable attention mechanism can be described with following operations:
\begin{equation*}
\begin{aligned}
R &= 
\left[
\begin{array}{c|c|c} 
R_1 & R_2 & R_3
\end{array}
\right] 
\\
M &= 
\left[
\begin{array}{c|c|c} 
M_1 & M_2 & M_3
\end{array}
\right]
\\
score &= R \cdot M \\
\alpha_{*} &= softmax(score) \\
O &= \sum_{n=1}^{3}\alpha_{n} * V_n
\end{aligned}
\end{equation*}
where $score$ is a one-by-three matrix and $V_n$ is GRU$_n$'s output. The matrix $R$ is an one-by-$r$ non-negative matrix and is the additional parameter matrix. $R_1$, $R_2$ and $R_3$ are partitioned matrices of $R$. Each represents the importance of the output of the GRU layer with the identical subscript ID. Matrix $M$, which is a $r$-by-three binary matrix, summarize each partitioned matrix by summing up its elements. This matrix is fixed once we set the column sizes for all $R_n$ where $n \in \{1,2,3\}$. If $R_n$ contains more elements than others, it is more likely that GRU$_n$’s output receives higher attention than others. Thus, we view matrix $R$ as the matrix of representatives and matrix $M$ as the matrix of tally. By setting the column size of $R$’s partitioned matrices, we can gently encourage the mechanism to assign extra attention on layers which are important in our prior knowledge. For example, if we think GRU$_1$ should receive more attention, we can define the column sizes for $R_1$, $R_2$ and $R_3$ to be 2, 1 and 1, respectively. The $R$ and $M$ should be the following format:
\begin{align*}
R &=
\left[
\begin{array}{cc|c|c}
ele_{1} & ele_{2} & ele_{3} & ele_{4}
\end{array}
\right]
\\
M &=
\left[
\begin{array}{c|c|c}
1 & 0 & 0\\
1 & 0 & 0\\
0 & 1 & 0\\
0 & 0 & 1\\
\end{array}
\right]
\end{align*}
where $ele_h$ is a scaler for all $h \in\{1,2,3,4\}$. We name this mechanism as the learnable attention mechanism (LAM).

\section{Data}
\label{sec: data}
The data comes from a long-term behavioral research project that uses internet-based social interactions as a tool to enhance seniors’ cognitive reserve. This project, titled as \textit{I-CONECT}, conducted at Oregon health \& Science University (OHSU), University of Michigan, and Wayne State University. Socially isolated older adults with above 80 years old are mainly recruited from the local Meals on Wheels program in Portland, Oregon and in Detroit, Michigan (with recruitment of African American subjects). Conversations are semi-structured, in which participants freely talk about a predefined topic, i.e. picnic, summer time, swimming and so on, with a moderator online. The corpus includes 30-minute recordings of  61 older adults (29 diagnosed with MCI,  25 normal controls, and 7 without clinical diagnosis) along with their professionally annotated transcriptions.

\subsection{Preprocess}
As our target speakers are seniors, we remove moderators' utterances based on the given speaker labels of utterances in the manual transcription. We extract word-level timestamps using a force-alignment algorithm available in \textit{Gentle}\footnote{https://github.com/lowerquality/gentle}, which is an open-source software, for each senior’s utterance. We segment long utterances into multiple pieces that are less than 7 seconds long by utilizing the word-level timestamps. Finally, we removed all utterances that are less than 3 seconds.

\subsection{Data Splitting}
\label{sce:data_split}
For both validation and testing sets, we randomly select 2 MCI and 2 healthy participants from both genders and leave 53 participants for the  training set. With 14 hours of transcribed  speech, this splitting leaves about $12.7$ hours of audio recordings for the training purpose. The total recording duration in the validation set and testing set are $0.66$ and $0.56$ hours, respectively. We use the validation set to select hyperparameters (i.e., learning rate) and the testing set is only used for assessing the model performance.

\section{Attention Over GRU Outputs}
\label{sec: exp_atten}
Our base model is an open-sourced DS2 model\footnote{https://github.com/PaddlePaddle/DeepSpeech}, which is trained on Baidu’s 8000 hours internal data, as well as the corresponding n-gram language model. The language model is fixed throughout all experiments. We have two base lines. We test the original model on our testing set. Also, we tune the entire model with our training set for 40 epochs and evaluate its performance. The tuned model's nickname is Plain WTL model (PTM).   

\subsection{Manual Attention Layer}
\label{sec: exp_manual_atten}
We use the MAM at the AttenL and define the basic attention unit to be $1/6$. All $\alpha$s must be an integral multiple of the unit. We use \emph{M-$\alpha_{1}$-$\alpha_{2}$-$\alpha_{3}$} to present the setting of attention in an experiment. For example, if we assign all attention to GRU$_1$, the attention setting is \emph{M-6/6-0/6-0/6}. In train process, we fine tune the modified model for 40 epochs.

\begin{figure}[htb]
    \centering
    \includegraphics[width=\linewidth]{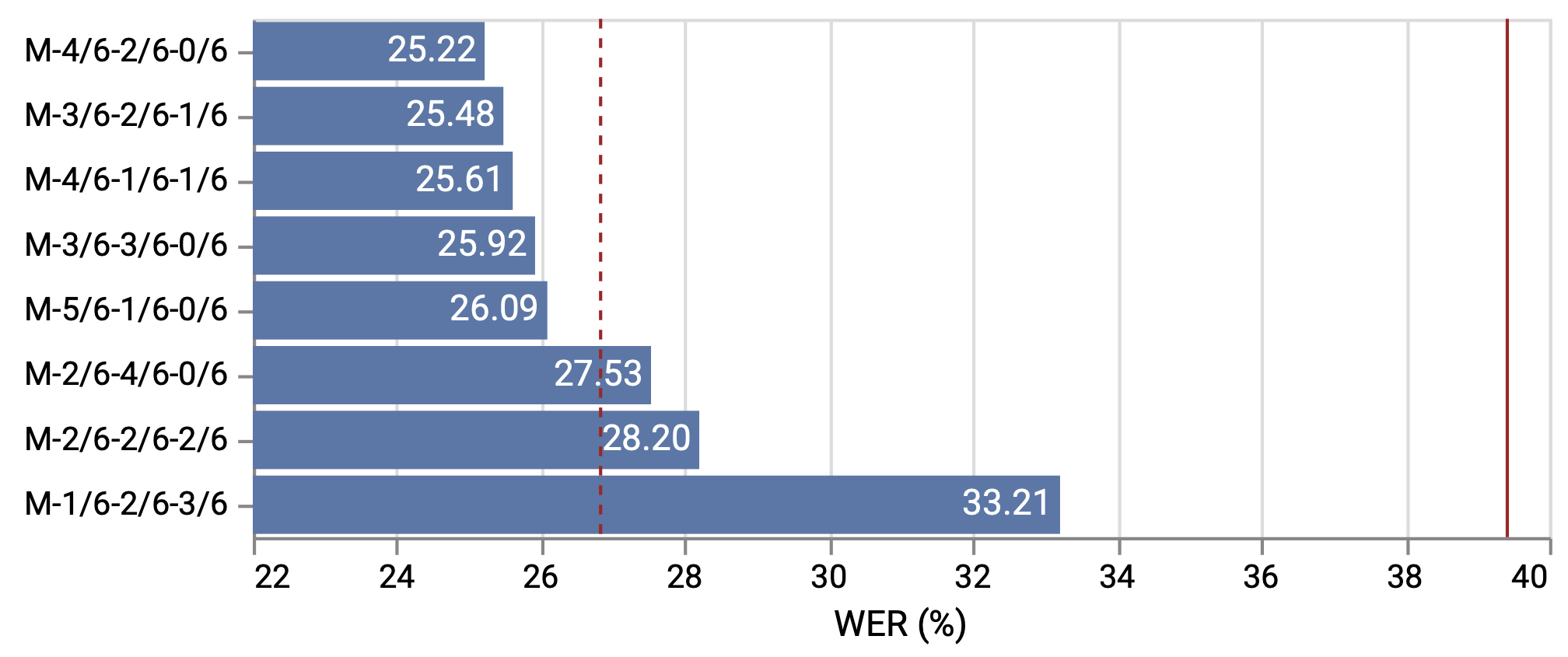}
    \caption{Model performance for manual attention settings. The red dotted line is WER of PTM ($26.8\%$). The red solid line is WER of the base model ($39.42\%$).}
    \label{fig:manual_scaling_results}
\end{figure}

In Fig~\ref{fig:manual_scaling_results}, the PTM completely outperforms the base mode on senior domain. The top 5 settings, where we assign more than half attention to GRU$_1$’s output, outperform PTM. The \emph{M-$4/6$-$2/6$-$0/6$} achieves $1.58\%$ absolute improvements over the PTM. Since we use the pre-trained Linear$_1$, who used to receive GRU$_1$’s output only, GRU$_1$ is naturally strongly related to Linear$_1$. On the other hand, unreasonable settings, assigning small attention to the output of GRU$_1$, perform worse than PTM. This experiment brings us confidence on utilizing intermediate information.

\subsection{Learnable Attention Layer}
\label{sec: exp_learnable_atten}

We adopt the LAM at the AttenL to learn the attention from the target dataset. We use \emph{L-$r_1$-$r_2$-$r_3$} to specify the column sizes of partitioned matrices in $R$. We evaluate four settings: \emph{L-1-1-1}, \emph{L-4-1-1}, \emph{L-3-2-1} and  \emph{L-5-4-0}. We use the first one to evaluate LAM’s learning ability. The other settings are designed to assign more attention to GRU$_1$’s output. All experiments first train the AttenL for $5$ epochs while freezing other layers. Then, we reverse the freezing status and train the model for $40$ epochs. 
We try each setting for $5$ times to evaluate the influence of randomly initialization on additional parameters.

\begin{figure}[htb]
    \centering
    \includegraphics[width=\linewidth]{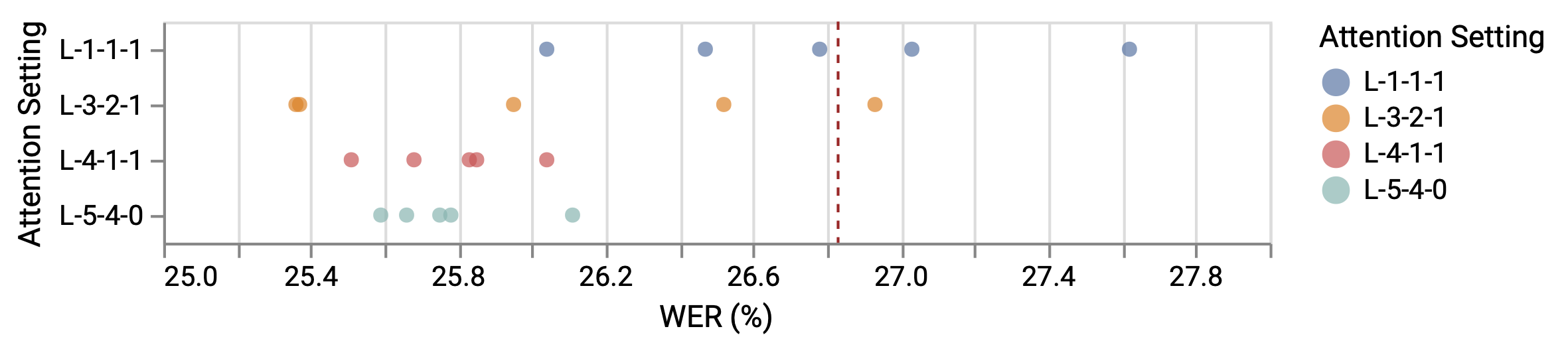}
    \caption{Performance on learnable attention settings. The red dotted line is WER of PTM.}
    \label{fig:atten}
\end{figure}

In Fig~\ref{fig:atten}, random initialization dramatically influences \emph{L-1-1-1}’s final outcome. On the contrary, we achieve more stable performance when applying prior knowledge through the column setting. This proves that setting the column sizes, based on prior knowledge, positively influences a tuned model's performance. Another evidence comes from the comparison between \emph{L-4-1-1} and \emph{L-3-2-1}. Although the total column size of $R$s are the same, models with \emph{L-4-1-1} perform more stable than the other. Both \emph{L-4-1-1} and \emph{L-5-4-0} outperform PTM. Our results are marginally worse than the optimal WER in Section~\ref{sec: exp_manual_atten}, but we cannot exclude the negative influence from the small training set. Moreover, LAM can be transformed to conditional-dependent form, which is a flexibility that MAM does not have. 

\section{Conclusion And Future Work}
\label{sec:conclusion}
We propose a conditional-independent attention mechanism to leverage a pre-trained model's intermediate information for model adaptation on the senior domain. We experimentally identify the domain mismatch between the pre-trained DS2 model and seniors and the benefit of applying TL. Our method, which stands on the shoulder of TL, can further reduce the mismatch. Also, our experiments support that guild the training direction with prior knowledge reduces the negative influence caused by random initialization. We will analysis how the size of training data influences the performance of learnable attention mechanism.

\section{Acknowledgements}
This work was supported by Oregon Roybal Center for Aging and Technology Pilot Program award P30 AG008017-30 in addition to NIH-NIA  awards R01-AG051628, and R01-AG056102.


\vfill
\pagebreak

\bibliographystyle{IEEEbib}
\bibliography{Template}

\end{document}